\documentclass[preprint]{iucr}              
\usepackage{amsmath,bm,diagbox}
\usepackage{algorithm}
\usepackage{algpseudocode}

\DeclareMathOperator*{\argmax}{arg\,max}

     \journalcode{J}              

\begin{document}                  



\title{Quantitative Selection of Sample Structures in Small-Angle Scattering Using Bayesian Methods}


\author[a]{Yui}{Hayashi}
\author[a]{Shun}{Katakami}
\author[b]{Shigeo}{Kuwamoto}
\author[c]{Kenji}{Nagata}
\author[d]{Masaichiro}{Mizumaki}
\cauthor[a]{Masato}{Okada}{*okada@edu.k.u-tokyo.ac.jp}

\aff[a]{Graduate School of Frontier Sciences, The University of Tokyo, Kashiwa, Chiba 277-8561, Japan}
\aff[b]{Japan Synchrotron Radiation Research Institute, Sayo, Hyogo 679-5198, Japan}
\aff[c]{National Institute for Materials Science, Tsukuba, Ibaraki 305-0047, Japan}
\aff[d]{Facalty of Advanced Science and Technology, Kumamoto University, Kumamoto 860-8555, Japan}








\maketitle                        
\begin{synopsis}
    We proposed a Bayesian method for quantitatively selecting a mathematical model of a sample for small-angle scattering, and evaluated its performance through numerical experiments on artificial data of a sample containing a mixture of multiple spherical particles.
\end{synopsis}

\begin{abstract}
    Small-angle scattering (SAS) is a key experimental technique for analyzing nano-scale structures in various materials.
    In SAS data analysis, selecting an appropriate mathematical model for the scattering intensity  is critical, as it generates a hypothesis of the structure of the experimental sample. 
    Traditional model selection methods either rely on qualitative approaches or are prone to overfitting.
    This paper introduces an analytical method that applies Bayesian model selection to SAS measurement data, enabling a quantitative evaluation of the validity of mathematical models.
    We assess the performance of our method through numerical experiments using artificial data for multicomponent spherical materials, demonstrating that our proposed method analysis approach yields highly accurate and interpretable results.
    We also discuss the ability of our method to analyze a range of mixing ratios and particle size ratios for mixed components, along with its precision in model evaluation by the degree of fitting.
    Our proposed method effectively facilitates quantitative analysis of nano-scale sample structures in SAS, which has traditionally been challenging, and is expected to significantly contribute to advancements in a wide range of fields.
\end{abstract}


\section{Introduction}
In recent years, the analysis of nano-scale structures of materials has become increasingly important in advancing the development of new materials and understanding biological phenomena. 
Small-angle scattering (SAS) is a fundamental experimental method for analyzing such nano-scale structures. 
It involves irradiating substances with X-rays or neutron beams and analyzing the resulting scattering intensity data at small angles, typically 5 degrees or less (Guinier \& Fournet, 1955).
SAS is versatile and applicable to a wide array of heterogeneous materials including nanoparticles, polymers, soft materials, and fibers, and utilized across many fields including material science, chemistry, and biology.\par
SAS measurement data are expressed in terms of scattering intensity that corresponds to a scattering vector, a physical quantity representing the scattering angle.
Data analysis requires selection and parameter estimation of a mathematical model of the scattering intensity, that contains information about the structure of the specimen.
This selection process is critical as it involves assumptions about the structure of the specimen.\par
Traditionally, model selection in SAS data analysis has been performed by listing candidate models based on theoretical or empirical rules, conducting parameter fitting against the measurements, and comparing suitability using criteria such as $\chi$-squared error, among other criteria (Da Vela \& Svergun, 2020; Svergun et al., 1995; Kline, 2006; Schneidman-Duhovny et al., 2010; Breßler et al., 2015).
Alternatively, models may be chosen based on the general shape of the measurement data.
However, these methods each have drawbacks: the former risks overfitting, which can lead to an overestimation of the model's degrees of freedom (Rambo \& Tainer, 2013), while the latter yields only qualitative model selections.
Furthermore, quantitatively evaluating the reliability of the results is challenging with traditional methods.\par
In this study, we propose a novel framework for the SAS model selection that quantitatively assesses the validity of mathematical models that represent specimen structures in measurements.
This approach uses Bayesian model selection within the framework of Bayesian inference, a method increasingly applied to analysis of various types of physical experimental data (Nagata et al., 2012; Nagata et al., 2019; Rappel et al., 2020; Machida et al., 2021; Nagai et al., 2021; Kashiwamura et al., 2022; Katakami et al., 2022; Ueda et al., 2023).
In the context of SAS data, Bayesian inference has been used for grain size distribution (Asahara et al., 2021) and parameter estimates (Hayashi et al., 2023).
This method solves inverse problems by establishing the likelihood, which is the data generation model, and the prior distribution that corresponds to the prior knowledge about the target being estimated. The posterior distribution is then calculated according to the model and parameters with the acquired data using Bayes' theorem.
In our proposed method, the posterior probability of the data generation model is calculated under the measured data using the Exchange Monte Carlo method (Hukushima \& Nemoto, 1996), also known as Parallel Tempering, and then comparing the resulting value among the candidate models while concurrently obtaining Bayesian estimates of the model parameters.
Moreover, since the validity of the measured data model is obtained as a posterior probability, the reliability of the results can be quantified by comparing these probabilistic values.
In this study, we conducted numerical experiments to assess the effectiveness of our proposed method.
These experiments are based on synthetic data used to estimate the number of distinct components in a specimen, which was modeled as a mixture of two types of monodisperse spheres of varying radii, scattering length densities, and volume fractions.
This type of problem is challenging due to the risk of overfitting, as the candidate models have similar 
structures; however, the results demonstrate high accuracy, interpretability, and stability 
of our method, even in the presence of measurement noise.\par
The structure of this paper is as follows: we first formalize the proposed analytical method, then discribe the model of multicomponent monodisperse spheres used in our numerical experiments.
In Sect. 4, we detail the set-up and results of these experiments using the proposed method to 
estimate the number of mixed components in the synthetic data.
We then discuss the analytical capabilities of our method and the performance of the traditional method based on the degree of fitting. 
We conclude with implications and potential applications of our method.

\section{Formulation of the Proposed Framework}
In this section, we present a detailed formulation of our algorithm for selecting mathematical models for SAS specimens using Bayesian model selection.
The pseudocode for this algorithm is provided in Algorithm 1.

\subsection{Bayesian Model Selection}
The process of generating experimental measurement data is generally described by a probabilistic model that includes noise components.
The SAS measurement data consist of scattering intensities that correspond to the scattering vector.
As the scattering intensity is a measure of the number of incident photons on the detector, the scattering intensity values are assumed to follow a Poisson distribution (Durant et al., 2021, Katakami et al., 2022, Nagata et al., 2019, Straasø et al., 2013).
Let $I_K(q, \Theta)$ be the mathematical model of scattering intensity characterized by the parameter $K$ for sample parameters $\Theta$ and the scattering vector $q$. The likelihood, which is the probability of generating the measured value $y$ is then given by the equation:

\begin{eqnarray}
    p\left(y|q,\Theta, K \right) &=& \frac{I_K(q, \Theta)^y \exp\left(-I_K(q, \Theta)\right)}{y!}. \label{poisson_noise}
\end{eqnarray}

\noindent Assuming that the measurement data $\mathcal{D} = \left\{q_i, y_i\right\}_{i=1}^N$, which consist of $N$ data points, are samples from an independent and identically distributed population under the $K$ and $\Theta$, the likelihood is expressed by the equation:
\begin{equation}
    p(\mathcal{D}|\Theta, K) = \prod^N_{i=1} p(y_i|q_i,\Theta, K). \label{likelihood}
\end{equation}
Here, we introduce the Poisson cost function to transform the likelihood of the measured data in Eq. (\ref{likelihood}) as:

\begin{equation}
    E(\Theta, K) = \frac{1}{N} \sum^N_{i=1} \left\{I_K(q_i,\Theta) - y_i\log I_K(q_i,\Theta) + \sum^{y_i}_{j=1} \log j\right\}.  \label{poisson_cost}
\end{equation}
The likelihood is thus expressed as:

\begin{equation}
    p(\mathcal{D}|\Theta, K) = \exp\left(-N E(\Theta, K)\right).
\end{equation}

Let $\varphi(K)$ be the prior distribution of the parameter $K$ that characterizes the model, and let $\varphi(\Theta|K)$ be the prior distribution of the model parameters $\Theta$. 
Then, from Bayes' theorem, the posterior distribution of the parameters given the measurement data can be written as:

\begin{eqnarray}
    p(\Theta|\mathcal{D}, K) &=& \frac{p(\mathcal{D}|\Theta, K)\varphi(\Theta|K)\varphi(K)}{\int p(\mathcal{D}, \Theta, K) d\Theta} \\
    &=& \frac{\exp\left(-N E(\Theta, K)\right)\varphi(\Theta|K)}{Z(K)},
\end{eqnarray}
\begin{equation}
    Z(K) = \int \exp\left(-N E(\Theta, K)\right)\varphi(\Theta|K) d\Theta. \label{target_marginal_likelihood}
\end{equation}

\noindent where $Z(K)$ is the marginal likelihood, which corresponds to the normalization constant of the posterior parameters distribution.
Furthermore, the probability of model $K$ given the data $\mathcal{D}$, denoted as $p(K|\mathcal{D})$, is given by the equation:

\begin{eqnarray}
    p(K|\mathcal{D}) &=& \frac{\int p(\mathcal{D}, \Theta, K)d\Theta}{\sum_K \int p(\mathcal{D}, \Theta, K)d\Theta} \\
    &=& \frac{\int \exp(-NE(\Theta, K))\varphi(\Theta|K)\varphi(K)d\Theta}{\sum_K \int \exp(-NE(\Theta, K))\varphi(\Theta|K)\varphi(K)d\Theta} \\
    &=& \frac{\exp(-F(K))\varphi(K)}{\sum_K \exp(-F(K))\varphi(K)}, \label{model_likelihood}\\
    \nonumber\\
    F(K) &=& -\log Z(K)\label{free_energy}.
\end{eqnarray}

\noindent $F(K)$ is referred to as the Bayesian free energy, also known as the stochastic complexity.
The posterior probability of the model, $p(K|\mathcal{D})$, can be rephrased as the validity of model $K$ for the measurement data $\mathcal{D}$.
In other words, calculating and comparing the value of $p(K|\mathcal{D})$ for all candidate models $\{K\}$ thus enables quantitative model selection.
Note that in Bayesian model selection, the parameter $K$ does not need to explicitly appear within the mathematical model of the specimen.

\subsection{Calculation of Marginal Likelihood}
In our Bayesian model selection method, the Bayesian free energy $F(K)$ and the probability $p(K|\mathcal{D})$ are calculated and compared for all candidate models. 
This computation relies on determining the marginal likelihood $Z(K)$, as expressed in Eq. (\ref{target_marginal_likelihood}). 
The marginal likelihood generally involves multi-dimensional integration, which can be computationally intensive and unstable.
To address this challenge, our framework uses the REMC to calculate the marginal likelihood (Hukushima \& Nemoto, 1996). 
This method facilitates sampling from the desired probability distribution at multiple inverse temperatures, referred to as replicas, using the Markov Chain Monte Carlo method (MCMC) to strategically exchange states between adjacent inverse temperatures at arbitrary intervals, thus avoiding local minima.
To calculate the marginal likelihood using REMC, we establish a series of $L$ inverse temperatures $\{\beta_l\}_{i=1}^L$  that they satisfy the relation:

\begin{equation}
0 = \beta_1 < ... < \beta_L = 1. \label{beta_relation}
\end{equation}

\noindent Sampling from the joint probability distribution at each inverse temperature gives:

\begin{equation}
p(\Theta_1, ..., \Theta_{L}|\mathcal{D}, K, \beta_1, ..., \beta_L) = \prod^L_{l=1}p(\Theta_l|\mathcal{D}, K, \beta_l),
\end{equation}
where $\Theta_l$ denotes the model parameter at the $l$-th inverse temperature $\beta_l$. 
The posterior distribution $p(\Theta_l|\mathcal{D}, K, \beta_l)$ satisfies the following relation:

\begin{equation}
p(\Theta_l|\mathcal{D}, K, \beta_l) \propto \exp(-N \beta_l E(\Theta_l, K))\varphi(\Theta_l|K). \label{posterior_dist}
\end{equation}

\noindent These distributions are sampled using MCMC at each inverse temperature, as expressed in Eq. (\ref{posterior_dist}), and states at adjacent inverse temperatures are periodically exchanged with a probability that satisfies the detailed balance condition. 
The probability of exchanging the $l$-th and $(l+1)$-th states, $p(\Theta_l \leftrightarrow \Theta_{l+1})$, is:

\begin{eqnarray}
    p(\Theta_l \leftrightarrow \Theta_{l+1}) &=& {\rm min} \left[1,\ \frac{p(\Theta_{l+1} | \mathcal{D}, K, \beta_l)p(\Theta_{l} | \mathcal{D}, K, \beta_{l+1})}{p(\Theta_{l} | \mathcal{D}, K, \beta_l)p(\Theta_{l+1} | \mathcal{D}, K, \beta_{l+1})} \right] \\
    &=& {\rm min} \left[1,\ \exp(N(\beta_{l+1} - \beta_{l})(E(\Theta_{l+1}, K) - E(\Theta_{l}, K))) \right].
\end{eqnarray}

\indent The marginal likelihood expressed in Eq. (\ref{target_marginal_likelihood}) can be efficiently determined using samples from various inverse temperatures sampled by REMC. 
The marginal likelihood $Z(K, \beta)$ at inverse temperature $\beta$ is expressed as:

\begin{equation}
Z(K, \beta) = \int \exp\left(-N \beta E(\Theta, K)\right)\varphi(\Theta|K) d\Theta.
\end{equation}

\noindent In this case, the target marginal likelihood expressed in Eq. (\ref{target_marginal_likelihood}) is equivalent to $Z(K, \beta=1)$. 
Using the relation in Eq. (\ref{beta_relation}), $Z(K, \beta=1)$ can be expressed as follows:

\begin{eqnarray}
Z(K, \beta=1) &=& \frac{Z(K, \beta_L)}{Z(K, \beta_{L-1})} \times \cdots \times \frac{Z(K, \beta_{2})}{Z(K, \beta_{1})} \\
&=& \prod ^{L-1} _{l=1} \frac{Z(K, \beta_{l+1})}{Z(K, \beta_{l})} \\
&=& \prod ^{L-1} _{l=1} \left\langle \exp\left(-N(\beta_{l+1} - \beta_{l}) E(\Theta_l, K) \right) \right\rangle _{p(\Theta_l|\mathcal{D}, K, \beta_l)}. \label{final_marginal_likelihood}
\end{eqnarray}

\noindent In Eq. (\ref{final_marginal_likelihood}), the symbol $\left\langle\cdot \right\rangle _{p(\Theta_l|\mathcal{D}, K, \beta_l)}$ 
denotes the expectation value with respect to $p(\Theta_l|\mathcal{D}, K, \beta_l)$. 
Computing Eq. (\ref{final_marginal_likelihood}) using sampling with REMC provides
the marginal likelihood expressed in Eq. (\ref{target_marginal_likelihood}). 
Once the marginal likelihood $Z(K)$ is determined, we can find the Bayesian free energy expressed in Eq. (\ref{free_energy}) and 
the posterior probability of model $K$ given the measurement data expressed in Eq. (\ref{model_likelihood}). 
For the numerical experiments presented here, we used the Metropolis method (Metropolis et al., 1953) for MCMC sampling of the posterior distributions at each inverse temperature, as expressed in Eq. (\ref{posterior_dist}).

\subsection{Estimation of Model Parameters}
During the marginal likelihood calculation the posterior distribution of $p(\Theta_L|\mathcal{D}, K, \beta_L=1)$ is obtained, which simply represents the Bayesian estimate of the model parameters (Hayashi et al., 2023). 
Therefore, the parameter estimation is conducted simultaneously as the Bayesian model selection performed.
Moreover, since the posterior distribution is sampled using REMC sampling, it can provide a global parameter estimate solution. 
Additionally, the reliability of the estimation can be assessed from the statistical properties of the sampled posterior distribution.\par
In Bayesian estimation, the Maximum A Posteriori (MAP) solution provides a point estimate of the parameters.
The MAP solution $\Theta_{\rm MAP}$ for the parameters of model $K$ is expressed by this equation from Eq. (\ref{posterior_dist}):

\begin{equation}
\Theta_{\rm MAP} = \argmax_{\Theta_L} \exp\left( -N E(\Theta_L,K) \right) \varphi(\Theta_L|K). \label{MAP_solution}
\end{equation}

\begin{algorithm}
    \textbf{Algorithm 1 :}
    \text{Proposed framework for quantitative selection of specimen model.}
    \begin{algorithmic}[1]
        \Require The measured data, $\mathcal{D} = \{q_i, y_i\}_{i=1} ^N$. The number of replicas, $L$. The inverse temperature, $\{\beta_l\}_{l=1}^L$ where $0 = \beta_1 < \cdots < \beta_L = 1$.
        The Burn-In, $\mathcal{T}_0$. The number of samples, $\mathcal{T}_1$. The step size for the Metropolis algorithm, $\{\epsilon_l\}_{l=1}^L$. 
        Candidate Models, $\{K_m\}_{m=1}^M$.The prior distribution of Models, $\varphi(K)$. 
        The prior distribution of parameters, $\varphi(\Theta|K)$.
        \Ensure $\forall l$ $\in \{1, \cdots, L\}$, $\forall \tau \in \{\mathcal{T}_0+1, \cdots,\mathcal{T}_1\} $, $\Theta_l^\tau \sim p(\Theta_l|\mathcal{D}, \beta_l)$. $Z(K)$, $F(K)$, $p(K|\mathcal{D})$.
        \For {$m \in \{1, \cdots, M\}$}
        \State Initialize array of sampled parameters, $\Psi = \{\}$.
        \For{$l \in \{1, \cdots, L \}$}
        \State $\Theta_l^0 \sim \varphi(\Theta|K_m)$
        \EndFor
        \For{$\tau \in \{1, \cdots, \mathcal{T}_0 + \mathcal{T}_1\}$}
        \For{$l \in \{1, \cdots, L \}$}
        \State Propose the following state, $\Theta'$ = $\Theta_l^{\tau-1}$ + $\epsilon \times \text{Uniform}(-1, 1)$.
        \State Calculate the acceptance ratio, $\alpha = {p(\Theta'|\mathcal{D}, K_m, \beta_l)}/{p(\Theta_l^{\tau-1}|\mathcal{D}, K_m, \beta_l)}$.
        \If{$\text{Uniform}(0, 1) < \alpha$}
        \State $\Theta_l^{\tau}$ = $\Theta'$
        \Else
        \State $\Theta_l^{\tau}$ = $\Theta_l^{\tau-1}$
        \EndIf
        \EndFor
        \For{$l \in \{1, \cdots ,L-1 \}$}
        \State Calculate the probability of exchanging states, $p(\Theta_l^\tau \leftrightarrow \Theta_{l+1}^\tau)$.
        \If{$\text{Uniform}(0, 1) < p(\Theta_l^\tau \leftrightarrow \Theta_{l+1}^\tau)$}
        \State Swap the $\Theta_l^\tau$ for the $\Theta_{l+1}^\tau$.
        \EndIf
        \EndFor
        \If{$\tau > \mathcal{T}_0$}
        \State Append the $\{\Theta_l^\tau\}_{l=1}^L$ to the $\Psi$.
        \EndIf
        \EndFor
        \State Calculate the marginal likelihood $Z(K_m)$ and the Bayesian free energy $F(K_m)$.
        \EndFor
        \State Calculate the likelihood of the model against the measured data $\{p(K_m|\mathcal{D})\}_{m=1}^M$.
    \end{algorithmic}
\end{algorithm}

\section{Formulation of a Multicomponent Monodisperse Spheres Model}
In this section, we describe a model for the scattering intensity of a dilute sample comprised of multicomponent monodisperse spheres (Guinier \& Fournet, 1955; Hashimoto, 2022).
This model serves as the basis for evaluating the performance of the proposed method. \par
Let $\bm{e}_i$ and $\bm{e}_s$ represent the unit vectors in the direction of the wave number vector of the incident and scattered beams, respectively. 
If $\bm{e}_i$ and $\bm{e}_s$ form an angle $2\alpha$, and the wavelength of the beam is $\lambda$, then the scattering vector $\bm{q}$ is given by:

\begin{equation}
    \bm{q} = \frac{4\pi \sin \alpha}{\lambda}\left(\bm{e}_s - \bm{e}_i\right).
\end{equation}
Assuming isotropic scattering, we consider the magnitude $q$ of the scattering vector. \par
The monodisperse spheres are spherical particles of uniform radius. 
The scattering intensity $I(q, \theta)$ of a specimen composed of sufficiently dilute monodisperse spheres of a single type for the scattering vector $q$ is given by:

\begin{equation}
    I(q, \theta) = SV\left(\frac{(\sin(qR) - qR\cos(qR))}{(qR)^3}\right)^2 + B,  \label{mono1}
\end{equation}
where $V = \frac{4}{3}\pi R^3$.
If the difference in scattering length density difference between the solute and solvent of the specimen is $\Delta \rho$ and the volume fraction is $\phi$, then $S = (3 \Delta \rho)^2 \phi$. The parameters $\theta$ of this model are the particle size $R$, the scale $S$, and the background $B$.\par
To formulate the scattering intensity of a specimen composed of $K$ types of monodisperse spheres, we assume a dilute system and denote the particle size of the $k$-th component in the sample as $R_k$ and the scale as $S_k$.
The scattering intensity of a sample composed of $K$ types of monodisperse spheres is then given by:

\begin{equation}
    I_K(q, \Theta) = \sum_{k=1}^{K} S_kV_k\left(\frac{(\sin(qR_k) - qR_k\cos(qR_k))}{(qR_k)^3}\right)^2 + B  \label{monoK}
\end{equation}
where we assume that $V_k = \frac{4}{3}\pi R_k^3$. 
The model parameters $\Theta$ for the scattering intensity $I_K(\cdot)$ are $\Theta = \left\{\{R_k, S_k\}_{k=1}^K, B \right\}$.

\section{Numerical Experiments}
Here, we present numerical experiments to evaluate the model selection among models with $K$ ranging from $1$ to $4$ components to demonstrate  the capabilities of the proposed framework.
We apply the framework to synthetic data generated to represent a system with two types ($K=2$) of monodisperse spheres, as described by Eq. (\ref{monoK}). 
This experimental design poses a challenge for discerning the true structure of the specimen; despite the simple structure, the similarity of the candidate models increases the risk of overfitting. \par
In typical SAS experiments, the scale parameter $S_k$ in Eq. (\ref{monoK}) tends to be small. 
Therefore, we normalize the scale parameter $S_k$ as:

\begin{equation}
\bar{S}_k = S_k \times 10^8. \label{scale_norm}
\end{equation}

\noindent We accordingly refer to the model parameters as $\Theta = \left\{\{R_k, \bar{S}_k\}_{k=1}^K, B \right\}$.\par
The numerical experiments reported in this section were conducted with a burn-in period of $10^5$ and a sample size of $10^5$ for the REMC. 
We set the number of replicas for REMC, the values of inverse temperature, and the step size of the Metropolis method, taking into consideration the state exchange rate and the acceptance rate.

\subsection{Generation of Synthetic Data}
This scattering intensity in SAS experiments, which is typically recorded as count data, is subject to be Poisson noise, as described by Eq. (\ref{poisson_noise}). 
We therefore generated synthetic data $\mathcal{D}$ using the procedure:

\begin{enumerate}
\item Set the number of data points to $N=400$, and define the scattering vectors at the $N$ equally spaced points within
the interval $[0.1, 3]$ to obtain $\{q_i\}_{i=1}^{N=400}$ [nm$^{-1}$].
\item Assume $K=2$ and set the true model parameters $\Theta$ to $\Theta^*$.
\item Calculate the scattering intensity at the scattering vectors $\{q_i\}_{i=1}^{N}$ obtained in step 1,
using the model in Eq. (\ref{monoK}) and $\Theta^*$. Introduce a pseudo-measurement time $T$ to adjust 
the noise intensity in the data, to obtain $\{I(q_i, \Theta^*, K)\times T\}_{i=1}^{N}$.
\item Generate measurement values $\{y_i\}_{i=1}^{N}$ as Poisson-distributed random numbers 
with means of $\{I(q_i, \Theta^*, K) \times T\}_{i=1}^{N}$ to create the synthetic dataset $\mathcal{D}=\{q_i, y_i\}_{i=1}^{N}$.
\end{enumerate}

\noindent In this section, we consider cases with pseudo-measurement times of $T=1$ and $T=0.1$. 
Generally, smaller values $T$ indicate the greater effects from measurement noise.

\subsection{Setting the Prior Distributions}
In the Bayesian model selection framework, prior knowledge concerning the parameters $\Theta$ and the model-characterizing parameter $K$ is set as their prior distributions. \par
In this numerical experiment, the prior distributions for the parameters $\Theta$ were set as Gamma distributions based on the pseudo-measurement time $T$ used during data generation, while the prior for $K$ was a discrete uniform distribution over the interval $[1, 4]$.

\begin{equation}
    \varphi(\Theta|K) = \varphi(B) \prod_{k=1}^K \varphi(R_k)\varphi(\bar{S}_k).
\end{equation}

\noindent
\begin{eqnarray}
    \varphi(R_k) &=& \text{Gamma}(R_k; \alpha = 1.2, \beta = 20) \\
    &=& \frac{\exp(-x/\beta)}{\beta^\alpha \cdot \Gamma (\alpha)} \cdot x^{\alpha - 1}, \\
    \varphi(\bar{S}_k) &=&  \begin{cases}
        \text{Gamma}(\bar{S}_k; 1.05, 300 \times T) & (\text{if $\{\bar{S}_k\}_{k=1}^K$ is in descending order}), \\
        0       & (\text{otherwise}), \label{prior_scale}
      \end{cases} \\
    \varphi(B) &=& \text{Gamma}(B; 1.05, 0.02 \times T),
\end{eqnarray}

\begin{equation}
    \varphi(K) = \text{DiscreteUniform}(K; 1, 4).
\end{equation}

\subsection{Results for Two-Component Monodisperse Spheres Based on Scale Ratio}
The ratio of scale parameters $S_1, S_2$ for spheres 1 and 2 during data generation, denoted as $r_S$, is defined as follows:

\begin{equation}
    r_S = \frac{\bar{S}_2}{\bar{S}_1}.
\end{equation}

\noindent Next, we present the analysis results from applying our proposed method to analyzing 6 types of data generated by varying the value of $r_S$ for pseudo measurement times of $T=1$ and $0.1$. 
Table \ref{tb:table1} displays the parameter values used for generating the synthetic data.

\begin{table}
    \caption{Parameter values used for data generation with varying $r_S$.}
    \label{tb:table1}
    \begin{center}
    \begin{tabular}{|c | c c|}
    \hline
    \diagbox{}{} & Sphere1  & Sphere2 \\
    \hline
    Radius $R$ (nm) & 2 & 10 \\

    Scale $\bar{S}$ & $250$ & $\{250, 100, 20, 0.5, 0.1, 0.05\}$ \\
    \hline
    Background $B$ (cm$^{-1}$) & $0.01$ & \\

    Pseudo measurement time $T$ & $\{1, 0.1\}$ & \\
    \hline
    \end{tabular}
    \end{center}
\end{table}

\begin{figure}
    \begin{center}
    \includegraphics[scale=0.7]{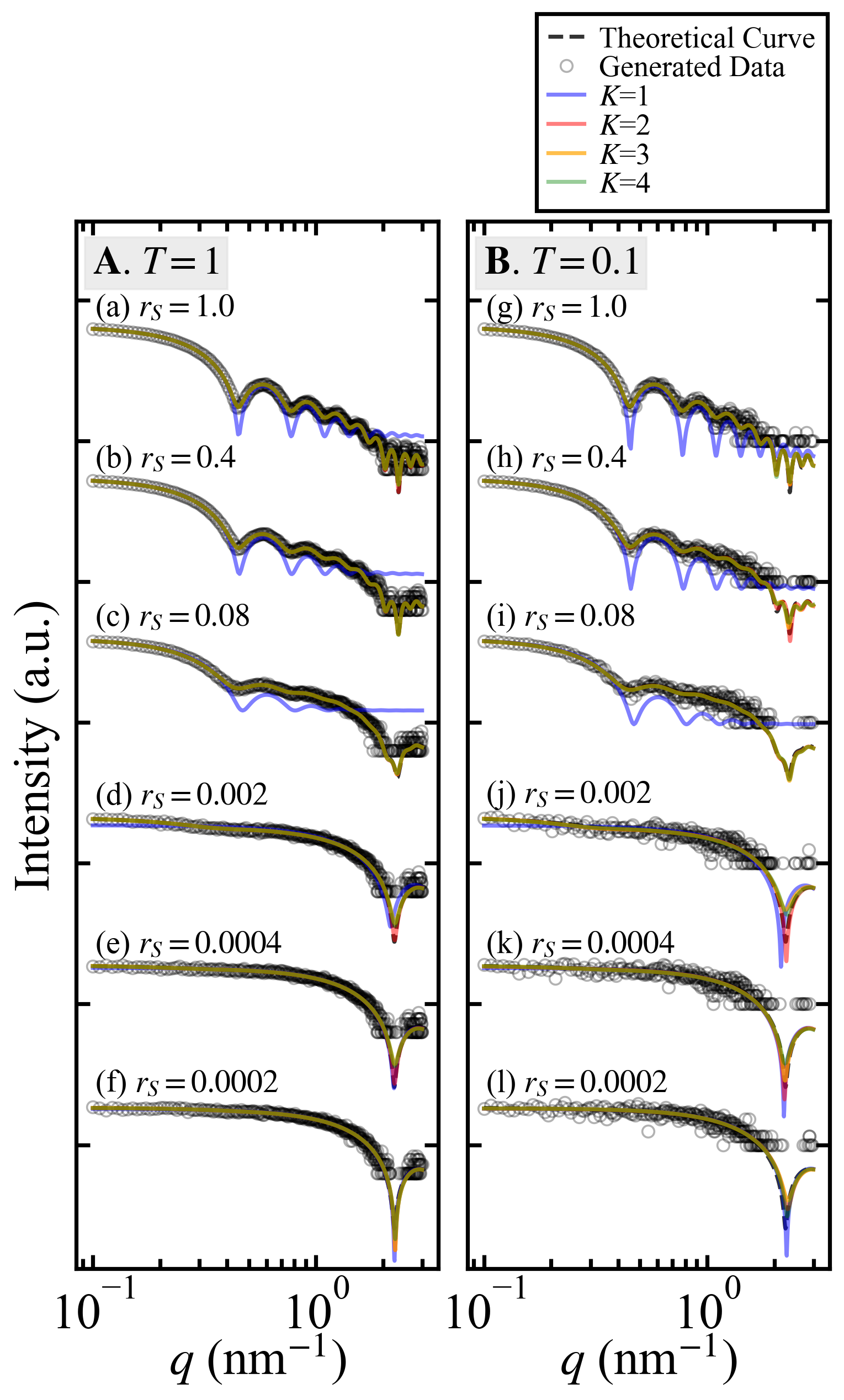}
    \caption{Fitting to synthetic data generated at various $r_s$ values with MAP solutions.\\
            Panels A and B show the cases for a pseudo-measurement time of $T=1$ and $T=0.1$, respectively. 
            In cases (a)--(f) and (g)--(l), the scale ratio $r_S$ is displayed in descending order for $T=1$ and $T=0.1$, respectively. 
            Black circles represent the generated data, the black dashed line represents the true scattering intensity curve, the solid blue, red, green, and orange lines represent the fitting curves using the MAP solution for $K=1$, $K=2$, $K=3$, and $K=4$, respectively.}
    \end{center}
    \label{fig:fig1}
\end{figure}

Figure \ref{fig:fig1} shows the fitting results for each model using the MAP solution for the synthetic data generated using the parameter values given in Table \ref{tb:table1}. 
In Fig. \ref{fig:fig1} (a) -- (c), (g) -- (i), it is apparent that the model with $K=1$ fails to accurately represent the data. 
However, we can also see that the fitting curves for models with $K=2 \-- 4$ are almost identical in shape. 
Furthermore, the data shown in Fig. \ref{fig:fig1} (d) -- (f), (j) -- (l) are difficult to distinguish from the well-known 
scattering data of a single type of monodisperse sphere ($K=1$), making it challenging to qualitatively compare the goodness of fit among the models with $K=1 \-- 4$.

\begin{figure}
    \begin{center}
        \includegraphics[scale=0.5]{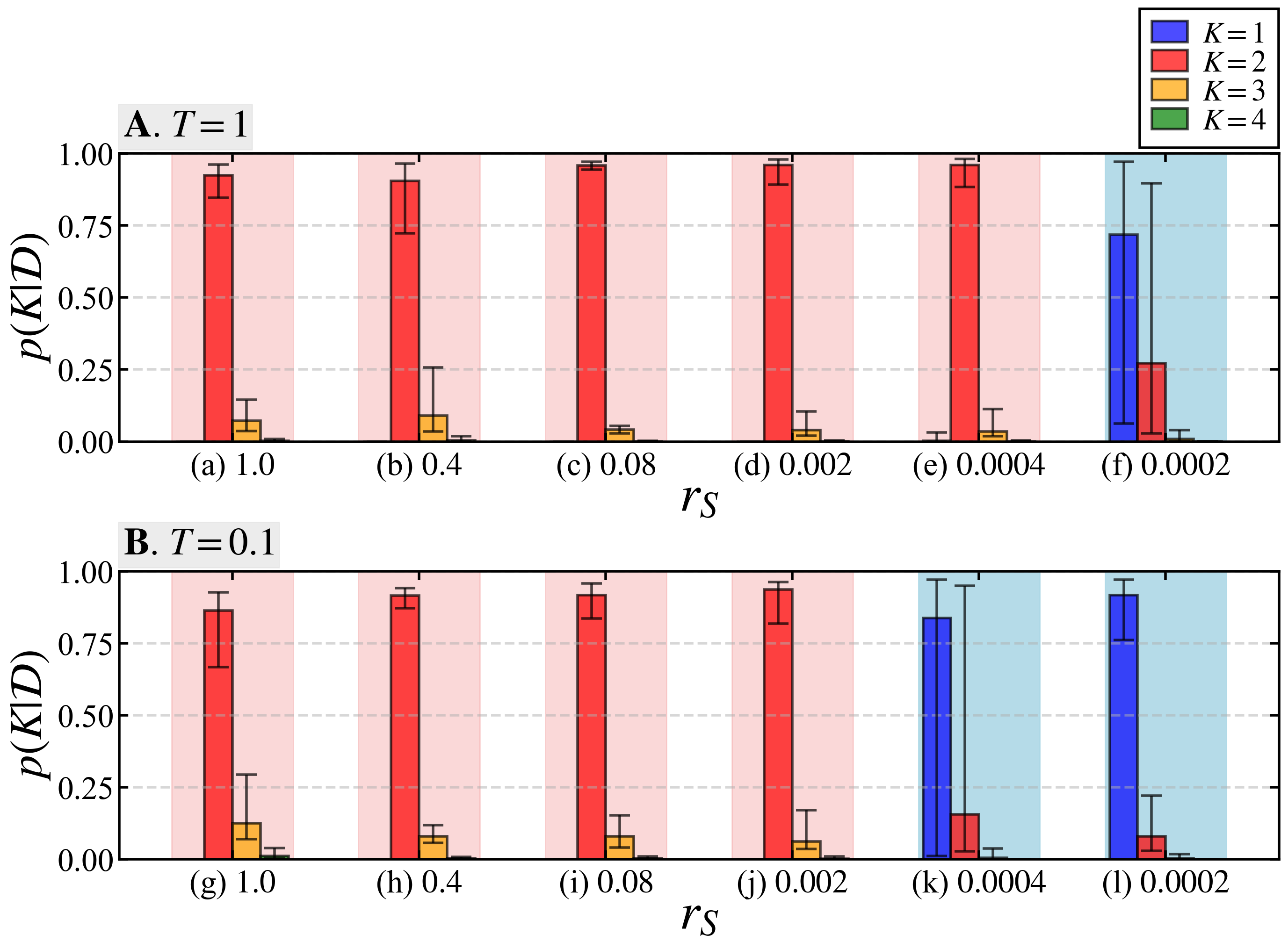}
        \caption{Results of Bayesian model selection among models $K=1 \-- 4$ for varying $r_S$ values.\\
        Panel A shows the posterior probability for each model using data generated with a pseudo-measurement time of $T=1$, and Panel B shows results for $T=0.1$. 
        In cases (a)--(f) and (g)--(l), the scale ratio $r_S$ is displayed in descending order for $T=1$ and $T=0.1$, respectively. 
        The height of each bar corresponds to the average values calculated for 10 datasets generated with different random seeds, with maximum and minimum values shown as error bars. 
        Areas highlighted in red indicates cases where, on average, the highest probability was found for the true model with $K=2$, while blue backgrounds indicate that models other than $K=2$ were associated with the highest probability on average.}
    \end{center}
    \label{fig:fig2}
\end{figure}

Figure \ref{fig:fig2} presents the Bayesian model selection results using our proposed framework. 
Figure \ref{fig:fig2}-A contains results for the case with $T=1$, and Figure \ref{fig:fig2}-B contains result with $T=0.1$, each showing the probability $p(K|\mathcal{D})$ of model $K$ based on the synthetic data $\mathcal{D}$ for each scale ratio 
$r_S$. 
Here, 10 datasets were created for each parameter value by varying the random seed during data generation, 
and the average value of $p(K|\mathcal{D})$ is indicated by the height of the bar graph, with error bars indicating the maximum and minimum values.
For the relatively large scale ratios $r_S$ in (a) -- (e) in Fig. \ref{fig:fig2}-A, the true model with $K=2$ has a high probability, while the average value of $p(K|\mathcal{D})$ is highest for $K=1$ in (f).
Conversely, in Fig. \ref{fig:fig2}-B, the true model with $K=2$ is associated with high probability in cases (g) -- (j), 
while in cases (k) and (l), $K=1$ is associated with the highest probability.

Table \ref{tb:table2} summarizes the number of times each model was found to have the highest probability in numerical experiments using the 10 separate datasets shown in Fig. \ref{fig:fig2}. 
For values of $r_S = 0.0004$ and above (Table \ref{tb:table2}-A), for $r_S = 0.002$ and above (Table \ref{tb:table2}-B), the model with $K=2$ was associated with the highest probability in all 10 datasets. 
This demonstrates the high accuracy of the proposed method and its robustness to measurement noise. 
In cases (f), (k), and (l), the model with $K=1$ was found to have the highest probability in nealy all of the 10 datasets. 
These results were used to inform a discussion on the suitable analysis range of $r_S$ using the proposed method, as addressed in the next section.

\begin{table}
    \caption{The number of times each model was associated with the highest probability in numerical experiments for 
            10 datasets generated with different random seeds at each $r_S$ values.\\
            In cases (a)--(f) and (g)--(l), the scale ratio $r_S$ is displayed in descending order for $T=1$ and $T=0.1$, respectively. 
            The most frequently counted case for each $r_S$ value is shown in bold.}
    \label{tb:table2}
    \begin{minipage}[t]{.45\textwidth}
      \begin{center}
        \text{A. $T=1$}\par
        \begin{tabular}{|l|cccc|}
            \hline
            \diagbox{$r_S$}{$K$} & 1 & 2 & 3 & 4 \\
            \hline
            (a) 1.0 & 0 & $\bm{10}$ & 0 & 0 \\
            (b) 0.4 & 0 & $\bm{10}$ & 0 & 0 \\
            (c) 0.08 & 0 & $\bm{10}$ & 0 & 0 \\
            (d) 0.002 & 0 & $\bm{10}$ & 0 & 0 \\
            (e) 0.0004 & 0 & $\bm{10}$ & 0 & 0 \\
            (f) 0.0002 & $\bm{8}$ & 2 & 0 & 0 \\
            \hline
        \end{tabular}
      \end{center}
      \label{table2_A}
    \end{minipage}
    \hfill
    \begin{minipage}[t]{.45\textwidth}
      \begin{center}
        \text{B. $T=0.1$}
        \begin{tabular}{|l|cccc|}
            \hline
            \diagbox{$r_S$}{$K$} & 1 & 2 & 3 & 4 \\
            \hline
            (g) 1.0 & 0 & $\bm{10}$ & 0 & 0 \\
            (h) 0.4 & 0 & $\bm{10}$ & 0 & 0 \\
            (i) 0.08 & 0 & $\bm{10}$ & 0 & 0 \\
            (j) 0.002 & 0 & $\bm{10}$ & 0 & 0 \\
            (k) 0.0004 & $\bm{9}$ & 1 & 0 & 0 \\
            (l) 0.0002 & $\bm{10}$ & 0 & 0 & 0 \\
            \hline
        \end{tabular}
      \end{center}
      \label{table2_B}
    \end{minipage}
  \end{table}

\subsection{Results for Two-Component Monodisperse Spheres Based on Radii Ratio}
During synthetic data generation, the ratio of the radii $R_1$ and $R_2$ of spheres 1 and 2, denoted as $r_R$, was defined as:

\begin{equation}
    r_R = \frac{R_1}{R_2}.
\end{equation}

\begin{figure}
    \begin{center}
        \includegraphics[scale=0.7]{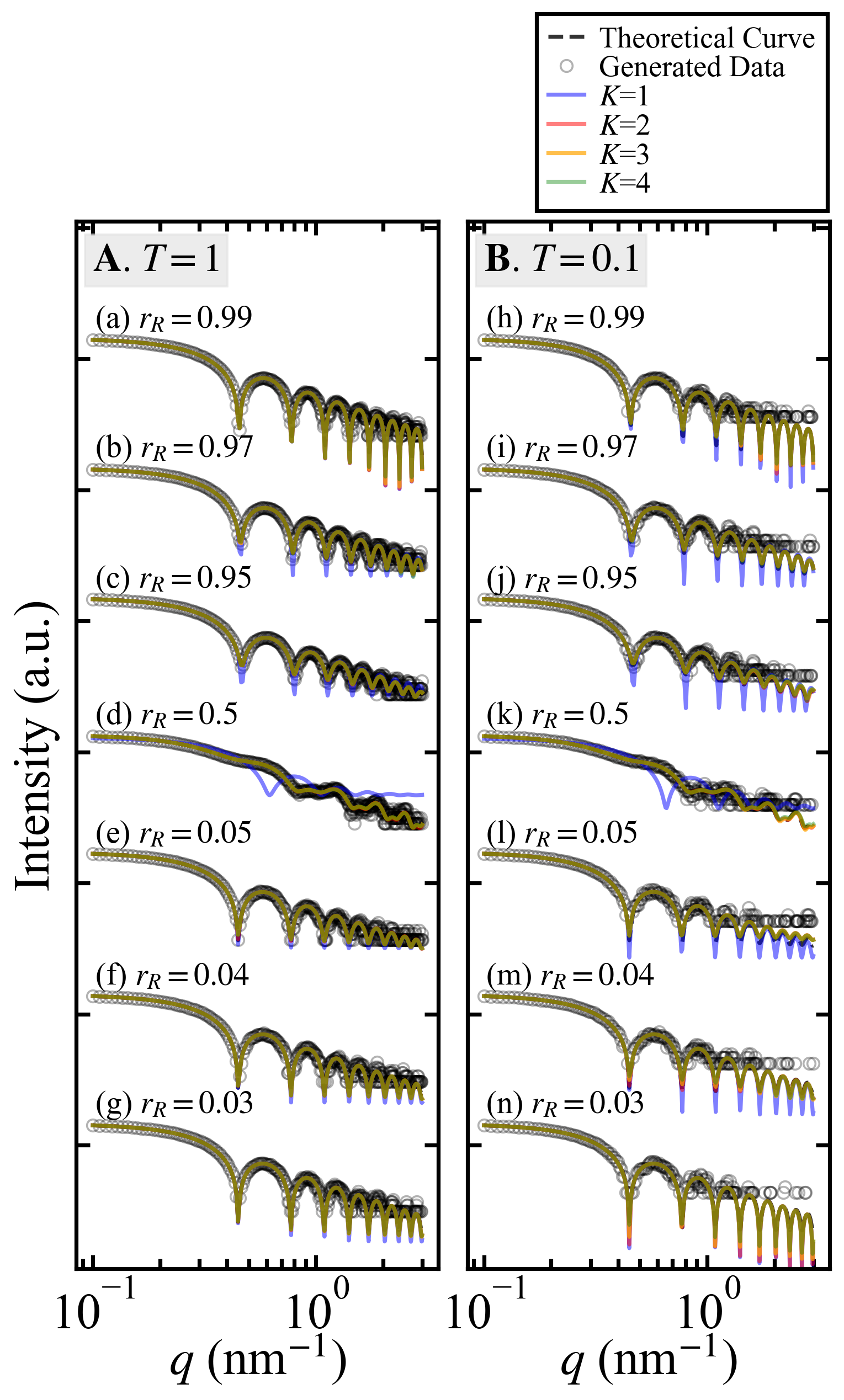}
        \caption{Fitting to synthetic data generated at various $r_R$ values with MAP solutions.\\
            Panels A and B show the cases for a pseudo-measurement time of $T=1$ and $T=0.1$, respectively.
            In cases (a)--(g) and (h)--(n), the radius ratio $r_R$ is displayed in descending order for $T=1$ and $T=0.1$, respectively. 
            The black dots represent the generated data, and the black dashed line represents the true scattering intensity curve, and the solid blue, red, green, and orange lines represent the fitting curves using the MAP solution for $K=1$, $K=2$, $K=3$, and $K=4$, respectively.}
    \end{center}
    \label{fig:fig3}
\end{figure}

\noindent In this set-up, we generated 7 types of data by varying the value of $r_R$ for pseudo measurement times of $T=1$ and $T=0.1$, respectively. 
The fitting results using the MAP solution with our proposed method to analyze these datasets are presented in Fig. \ref{fig:fig3}, while the parameter values used for data generation are given in Table \ref{tb:table3}.

\begin{table}
    \caption{Parameter values used for data generation when varying $r_R$.}
    \label{tb:table3}
    \begin{center}
    \begin{tabular}{|c | c c|}
    \hline
    \diagbox{}{} & Sphere1  & Sphere2 \\
    \hline
    Radius $R$ (nm) & $\{9.9, 9.7, 9.5, 0.5, 0.5, 0.4, 0.3\}$ & 10 \\

    Scale $\bar{S}$ & $250$ & $100$ \\
    \hline
    Background $B$ (cm$^{-1}$) & $0.01$ & \\

    Pseudo measurement time $T$ & $\{1, 0.1\}$ & \\
    \hline
    \end{tabular}
    \end{center}
\end{table}

\noindent Aside from the cases of $r_R=0.5$ in (d) and (k), the profiles of the data in Fig. \ref{fig:fig3} are very similar to those of a single monodisperse sphere, and the fitting curves for models $K=1$ to $K=4$ are nearly identical in shape. 
In contrast, the data for cases (d) and (k) with $r_R=0.5$ have a complex profile, and it can be seen that the model with $K=1$ can be seen to represent the data poorly.

\begin{figure}
    \begin{center}
        \includegraphics[scale=0.5]{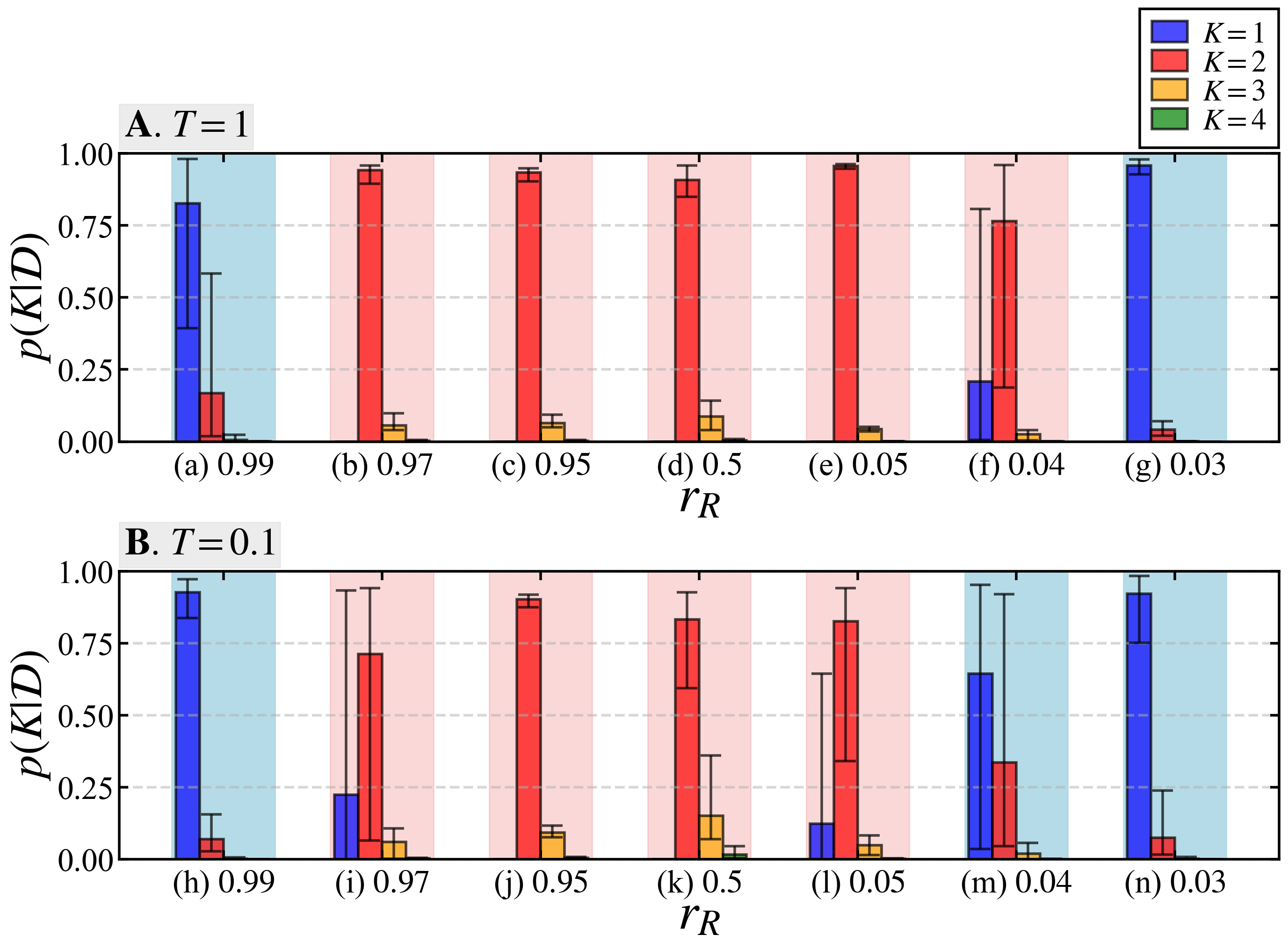}
        \caption{Results of Bayesian model selection among models $K=1$ -- $4$ for varying $r_R$ values.\\
                Panel A shows the posterior probability of each model using data generated with a pseudo-measurement time of $T=1$, and Panel B shows results for $T=0.1$.
                In cases (a)--(g) and (h)--(n), the radius ratio $r_R$ is displayed in descending order for $T=1$ and $T=0.1$, respectively. 
                The height of each bar corresponds to the average values calculated for 10 datasets generated with different the random seeds, with the maximum and minimum values shown as error bars.
                Areas highlighted in red indicates cases where the true model $K=2$ was most highly supported, while the blue backgrounds indicate that the likelihood of a model other than $K=2$ was the highest.}
    \end{center}
    \label{fig:fig4}
\end{figure}

Figure \ref{fig:fig4} displays the results of Bayesian model selection using synthetic data generated by varying the radius ratio $r_R$. 
10 datasets were created for each parameter value by varying the random seed during data generation, and the average value of $p(K|\mathcal{D})$ is indicated by the height of the bar graph, with the maximum and minimum values shown as error bars. 
Unlike the results for the variations in scale ratio shown in Fig. \ref{fig:fig2}, the model selection procedure fails not only at a radius ratio $r_R$ is close to 0, but also at values is close to 1, with $K=1$ being the most highly supported. 
Additionally, in the case of $r_R = 0.04$, the result for $T=1$ in case (f) supports the true model $K=2$, but for $T=0.1$ in case (m), the alternative model $K=1$ is most supported. 
However, in cases (b) -- (f) and (i) -- (l), the true model $K=2$ is associated with a high average probability (Fig. \ref{fig:fig4}).

Table \ref{tb:table4} presents the results of numerical experiments for the 10 separate datasets shown in Fig. \ref{fig:fig4}, summarizing the number of times each model $K=1$ -- $4$ was most highly supported. 
Near the analytical limits of the proposed method, there are cases where the supported model changes depending on the data, as shown in Table \ref{tb:table4} (a), (f), (i), (l), and (m). 

\begin{table}
    \caption{The number of times each model was most highly supported in numerical experiments for 10 datasets 
            generated by varying $r_R$ values.\\
            In cases (a)--(g) and (h)--(n), the radius ratio $r_R$ is displayed in descending order for $T=1$ and $T=0.1$, respectively. 
            The cases with the highest counts for each $r_R$ value are shown in bold.}
    \label{tb:table4}
    \begin{minipage}[t]{.45\textwidth}
      \begin{center}
        \text{A. $T=1$}\par
          \begin{tabular}{|l|cccc|}
            \hline
            \diagbox{$r_R$}{$K$} & 1 & 2 & 3 & 4 \\
            \hline
            (a) 0.99 & $\bm{9}$ & 1 & 0 & 0 \\
            (b) 0.97 & 0 & $\bm{10}$ & 0 & 0 \\
            (c) 0.95 & 0 & $\bm{10}$ & 0 & 0 \\
            (d) 0.5 & 0 & $\bm{10}$ & 0 & 0 \\
            (e) 0.05 & 0 & $\bm{10}$ & 0 & 0 \\
            (f) 0.04 & 1 & $\bm{9}$ & 0 & 0 \\
            (g) 0.03 & $\bm{10}$ & 0 & 0 & 0 \\
            \hline
        \end{tabular}
      \end{center}
      \label{table4_A}
    \end{minipage}
    \hfill
    \begin{minipage}[t]{.45\textwidth}
      \begin{center}
        \text{B. $T=0.1$}\par
          \begin{tabular}{|l|cccc|}
            \hline
            \diagbox{$r_R$}{$K$} & 1 & 2 & 3 & 4 \\
            \hline
            (h) 0.99 & $\bm{10}$ & 0 & 0 & 0 \\
            (i) 0.97 & 2 & $\bm{8}$ & 0 & 0 \\
            (j) 0.95 & 0 & $\bm{10}$ & 0 & 0 \\
            (k) 0.5 & 0 & $\bm{10}$ & 0 & 0 \\
            (l) 0.05 & 1 & $\bm{9}$ & 0 & 0 \\
            (m) 0.04 & $\bm{7}$ & 3 & 0 & 0 \\
            (n) 0.03 & $\bm{10}$ & 0 & 0 & 0 \\
            \hline
        \end{tabular}
      \end{center}
      \label{table4_B}
    \end{minipage}
  \end{table}

\section{Discussion}
In Sect. 4, we conducted numerical experiments to determine the number of components $K$ in two-component monodisperse sphere specimens using the proposed method through model selection applied to artificial measurement data. 
In this section, we discuss the analytical limits under the settings of this study concerning the scale ratio $r_S$ and radius ratio $r_R$ of the specimen's two components, as well as the conventional method of model selection based on the quality of fit often employed in SAS data analysis. 

\subsection{Limitations of the Proposed Method}
The experiments detailed in Sect. 4 expolored the selection of the number of components $K$ using the proposed method for two-component monodisperse spheres using the proposed Bayesian method. 
We observed certain analytical limitations for various values of the scale $r_S$ and radius ratio $r_R$.
In practical data analysis applications using the proposed method, it is advisable to conduct preliminary tests using synthetic data with noise intensity and anticipated parameter values similar to those of the measured data. 
This step can help ensure a more reliable analysis, as detailed below.\par
The scale parameter $S$ is a value that is multiplied by the square of the difference in scattering length density between the solvent and the specimen, as well as the volume fraction. 
This can cause $r_S$ to become extremely small when there is little difference in scattering length density between the solvent and a component of the specimen, or when there is a significant difference in the mixing ratio of the components. 
The results in Fig. \ref{fig:fig2} and Table \ref{tb:table2} for a pseudo-measurement time of $T=1$ (Panel A) indicates that the model selection favored non-true models at a scale ratio of $r_S = 0.0002$. 
Similarly, for $T=0.1$ (Panel B), non-true models were favored at scale ratios of $r_S = 0.0004$ and $r_S = 0.0002$, indicating that these cases exceed the analytical capabilities of the proposed method. 
These findings imply that within experimental parameters of this study, the proposed method reliably identifies the true model with a high probability for scale ratios up to $r_S=0.0004$ at $T=1$ and up to $r_S = 0.002$ at $T=0.1$.\par
In Sect. 4.2, we investigated the effect of varying the radius ratio $r_R$. 
When components of different radii are mixed, it is important to consider not only simple mixtures but also instances of aggregated specimens.
The findings shown in Fig. \ref{fig:fig4} and Table \ref{tb:table4} indicate that the proposed method reaches its analytical limits as $r_R$ approaches 1 and as it approaches 0. 
As $r_R$ nears 1, the scattering profiles of the two-component system become similar to that of a single-component system, leading to the selection of the single-component model ($K=1$). 
We identified an analytical limit at $r_R = 0.99$ for both $T=1$ and $T=0.1$. 
The results for $r_R=0.97$ show that at $T=0.1$, which has a higher noise intensity compared to $T=1$, the posterior probability of the single-component model ($K=1$) increases, resulting in an unstable analysis. 
Conversely, as $r_R$ approaches 0, the results $r_R = 0.03$ at $T=1$ and $r_R = 0.04$ and $r_R = 0.03$ at $T=0.1$, the single-component model ($K=1$) is associated with high probability, indicating an analytical limit. 
Overall, the proposed method demonstrates the ability to select the true model with high probability for 
radius ratios ranging from $r_R = 0.04$ to $0.99$ at $T=1$, and from $r_R = 0.05$ to $0.99$ at $T=0.1$.

\subsection{Model Selection Based on Cost Function Values}
Selecting an appropriate mathematical model is a critical step in SAS data analysis, yet it remains a challenge. 
Traditionally, model fitting involved comparing a set of candidate models using metrics such as the $\chi$-squared error to assist in model selection. 
However, the $\chi$-squared error assumes a Gaussian distribution of the measurement noise. 
In this study, we assume Poisson-distributed measurement noise, and therefore discuss the results of model selection by calculating and comparing the values of the Poisson cost function, as formulated in Eq. (\ref{poisson_cost}), across models $K=1 \-- 4$ using the measurement data and fitting curves derived from the MAP solutions.\par
Table \ref{tb:table5} reports the frequency of each model minimizing the Poisson cost value for models with $K=1$ to $K=4$. 
These results are based on the same datasets described in Sect. 4.4, which were generated by varying the random seed for each of the 6 distinct $r_S$ values determined by the parameters listed in Table \ref{tb:table1}, with 10 datasets produced for each $r_S$ value at $T=1$.

\begin{table}
    \caption{Results of model selection based on the Poisson cost function.\\
            The cases (a)--(f) correspond to the settings in Figs. 1 -- 3 of Sect. 4.4.
             The table indicates the number of times each model was found to have the lowest Poisson cost value for 10 datasets generated with different random seeds for each $r_S$ value at $T=1$. The most highly supported model, if unique, is shown in bold.}
    \label{tb:table5}
    \begin{center}
    \begin{tabular}{|l | c c c c|}
        \hline
        \diagbox{$r_S$}{$K$} & 1 & 2 & 3 & 4 \\
        \hline
        \hline
        (a) 1.0 & 0 & 4 & 2 & 4 \\
        (b) 0.4 & 0 & 2 & $\bm{6}$ & 2 \\
        (c) 0.08 & 0 & 3 & $\bm{7}$ & 0 \\
        (d) 0.002 & 0 & $\bm{6}$ & 3 & 1 \\
        (e) 0.0004 & 0 & 5 & 5 & 0 \\
        (f) 0.0002 & 1 & $\bm{5}$ & 4 & 0 \\
        \hline
    \end{tabular}
    \end{center}
\end{table}

The model selection results based on the Poisson cost function shown in Table \ref{tb:table5} indicate that for (a) and (e), there are two models most supported, suggesting that it is difficult to obtain reliable results.
In cases (b) and (c), the $K=3$ model is most supported, likely as a result of overfitting, and failing to select the correct model, $K=2$. 
Furthermore, when the true model $K=2$ was selected in cases (d) and (f), the $K=3$ model is supported similarly often. 
The above findings demonstrate that evaluating how well a model fits the measurement data based on residual values such as the Poisson cost and candidate model comparison dose not consistently result in selection of the true model.\par
In contrast the proposed method results in Table \ref{tb:table2}-A demonstrate that the correct model $K=2$ is supported in all 10 out of 10 cases from (a) to (e). 
Within the analyzable range described in the previous subsection, the model selection results are highly stably against data noise.

\section{Conclusions}
In this paper, we introduced Bayesian model selection framework for SAS data analysis that quantitatively evaluates model validity through posterior probabilities.
We conducted numerical experiments using synthetic data for a two-component system of monodisperse spheres to assess the performance of the proposed method. 
We identified the analytical limits of the proposed method, under the settings of this study, with respect to the scale and radius ratios of two-component spherical particles. 
Additionally, we compared its performance to traditional fit-based model selection methods based on the Poisson cost function. 
The numerical experiments and subsequent discussion revealed the range of parameters that can be analyzed using the proposed method. 
Within that range, our method provides stable and highly accurate model selection even for data with significant noise or situations in which qualitative model determination is challenging.
Furthermore, in comparison to the traditional method of selecting models based on fitting curves and data residuals, it was found that the proposed method offers greater accuracy and stability.\par
SAS is used to study specimens with a variety of structures other than spheres, including cylinders, core-shells, lamellae, and more. 
The proposed method should be applied other sample models to determining the feasibility of expanding the analysis beyond the case examined here to broader experimental settings. 
Future work can benefit from using the proposed method to conduct real data analysis, which is expected to yield new insights through our more efficient analysis approach.

\section{Funding information}
    This work was supported by JST CREST (Grant Nos. PMJCR1761 and JPMJCR1861) from the Japan Science and Technology Agency (JST) and JSPS KAKENHI Grant-in-Aid for Scientific Research(A) (No. 23H00486).

\end{document}